\documentclass[submission, Proceedings]{SciPost}

\usepackage{listings}
\usepackage{xcolor}
\usepackage{dsfont}
\usepackage{physics}
\usepackage{amsmath}
\usepackage{amssymb} 
\usepackage{float}

\newtheorem{definition}{Definition}[section]

\definecolor{codegreen}{rgb}{0,0.6,0}
\definecolor{codegray}{rgb}{0.5,0.5,0.5}
\definecolor{codepurple}{rgb}{0.58,0,0.82}
\definecolor{backcolour}{rgb}{0.95,0.95,0.92}

\lstdefinestyle{mystyle}{
    backgroundcolor=\color{backcolour},   
    commentstyle=\color{codegreen},
    keywordstyle=\color{magenta},
    numberstyle=\tiny\color{codegray},
    stringstyle=\color{codepurple},
    basicstyle=\ttfamily\footnotesize,
    breakatwhitespace=false,         
    breaklines=true,                 
    captionpos=b,                    
    keepspaces=true,                 
    numbers=left,                    
    numbersep=5pt,                  
    showspaces=false,                
    showstringspaces=false,
    showtabs=false,                  
    tabsize=2
}

\newcommand{\rdsymbol}{\texttt{RDSymbol} }
\newcommand{\rdbasis}{\texttt{RDBasis} }
\newcommand{\effectiveframe}{\texttt{EffectiveFrame} }
\newcommand{\block}{\texttt{Block} }

\begin{document}

\begin{center}{\Large \textbf{
SymPT: a comprehensive tool for automating effective Hamiltonian derivations}}\end{center}
\begin{center}
Giovanni Francesco Diotallevi\textsuperscript{1*},
Leander Reascos\textsuperscript{1*},
M\'onica Benito\textsuperscript{1}
\end{center}

\begin{center}
{\bf 1} Augsburg University$,$ Institute of Physics$,$ Universitätsstraße 1 (Physik Nord)$,$ 86159 Augsburg
\\
*francesco.diotallevi@uni-a.de, irving.reascos.valencia@uni-a.de
\end{center}

\def\thefootnote{*}\footnotetext{These authors contributed equally to this work.}\def\thefootnote{\arabic{footnote}}
\begin{center}
\today
\end{center}


\section*{Abstract}
The Schrieffer-Wolff transformation (SWT) is a foundational perturbative method for deriving effective Hamiltonians in quantum systems by systematically eliminating couplings between pairs of energy distant subspaces. Despite recent advancements, the implementation of SWTs for sufficiently complex systems remains computationally challenging and often requires extensive calculations that are prone to errors.
In this work, we introduce an analytical software tool, \href{https://github.com/qcode-uni-a/SymPT}{SymPT} (Symbolic Perturbation Theory), designed to automate the SWT and its extensions. Building on a universal framework developed in recent research, SymPT provides a systematic and generalizable solution for deriving the generator of the transformation, enabling accurate computation of effective Hamiltonians for arbitrary perturbative systems. The tool supports both time-independent and time-periodic Hamiltonians, extending beyond standard SWT to incorporate arbitrary coupling elimination, block-diagonalization and full-diagonalization routines, thus enabling precise handling of systems with intricate energy structures. SymPT is a free software available at \href{https://github.com/qcode-uni-a/SymPT}{\textbf{https://github.com/qcode-uni-a/SymPT}}. 

\newpage
\vspace{10pt}
\noindent\rule{\textwidth}{1pt}
\tableofcontents\thispagestyle{fancy}
\noindent\rule{\textwidth}{1pt}
\vspace{10pt}

\newpage
\section{Introduction}
\label{sec: Introduction}
The Schrieffer-Wolff transformation (SWT)~\cite{SW_original_paper_on_SW, SW_wikler_book, SW_TD_subharmonic_transition_and_block_siegert, Loss_DiVincenzo_SW} is a perturbative method used to simplify the analysis of quantum systems by mapping an otherwise complex Hamiltonian to an effective one that operates within a reduced subspace. This technique is particularly useful when dealing with systems with different energy scales, allowing for the removal of high-energy contributions that are irrelevant to the low-energy dynamics. By focusing on the interactions that govern the most significant physical behaviors, the SWT has become a valuable tool in fields such as condensed matter physics~\cite{SW_effective_hopping_in_silicene,SW_fermi_hubbard_model,SW_ising_quantum_criticality,SW_kondo_lattice,SW_mott_insulators,SW_superconductor_physics,SW_quantum_lattice_models,SW_superconductor_physics_1,SW_superconductive_physics_2,SW_XXZ_chain}, quantum information~\cite{SW_anomalous_zero_field_splitting, SW_flopping_mode_qubits, SW_quantum_algorithms_1, SW_recent_advances_in_hole_spin_qubits,SW_super_conducting_qubits} and quantum optics~\cite{Transmon_resonator_2,SW_dispersive_two_qubit_gates, SW_optical_lattice_interacting_bosons, SW_cavity_QED, SW_dissipative_bosonic_modes} where it aids in understanding phenomena like effective spin interactions and dispersive couplings in quantum circuits. Through its ability to isolate the essential dynamics of a system, the SWT plays a key role in simplifying both analytical and numerical approaches to quantum problems.

In a recent paper~\cite{universal_solution_paper}, we introduced a unified and systematic framework for the SWT, providing a closed-form solution for the transformation generator that addresses the shortcomings of previous methods. This framework, which is applicable to both time-independent and time-dependent systems, offers a general solution that depends solely on the perturbation being eliminated, overcoming the limitations of heuristic approaches and dimensional truncation. However, despite these theoretical advancements, the implementation of SWTs remains a complex task, particularly for systems with intricate Hamiltonians or those involving infinite-dimensional Hilbert spaces. The process of manually deriving the effective Hamiltonian can be cumbersome, requiring involved calculations that are both time-consuming and prone to error. 
Despite one notable implementation very well optimized to automate SWTs~\cite{pymablock}, no other comprehensive software tools currently exist that can automate this process without requiring Hilbert space truncation or the inclusion of additional information, outside of the system's Hamiltonian. 
Additionally, for time dependent systems a general-purpose tool capable of performing SWTs as well as its several extensions has not (to our knowledge) yet been released. This gap poses a significant barrier to the broader adoption and application of the SWT in both theoretical and practical research.

The objective of this paper is to address this gap by presenting \href{https://github.com/qcode-uni-a/SymPT}{SymPT}, an analytical software tool that automates the process of performing SWTs. This tool leverages the theoretical results derived in our previous work~\cite{universal_solution_paper} to provide a robust and efficient platform for computing effective Hamiltonians across a variety of quantum systems, enabling the perturbative treatment of Hamiltonians also at an operator level, thus without requiring any Hilbert space truncation. The software is designed to handle both time-independent and time-periodic perturbations, offering accurate results for systems that range from simple finite-dimensional models to more complex infinite-dimensional Hilbert spaces. Additionally, the tool is capable of extending beyond the conventional SWT, and other block-diagonalization routines, thereby broadening its applicability to a wide range of possible transformations.

\section{Methods}
\label{sec: Methods}
\subsection{The Schrieffer-Wolff transformation and its extensions} 
\label{sec: the schrieffer-wolff transformation}

As stated, the SWT is a perturbative approach used to derive effective Hamiltonians by systematically eliminating couplings between low- and high-energy subspaces of a given Hamiltonian, $\mathcal{H}$. Hamiltonians treatable with this approach are typically partitioned as 
\begin{align}
\mathcal{H} = \sum_{i=0} \mathcal{H}^{(i)} + \sum_{j=1} V^{(j)},\label{eq: decomposed hamiltonian}
\end{align} where $\mathcal{H}^{(0)}$ is the unperturbed component with a known spectrum of eigenstates. These eigenstates are divided into a low-energy subspace $\{\ket{L}\}$ and a high-energy subspace $\{\ket{H}\}$, connected by perturbative terms $V^{(j)}$. The SWT aims to block-diagonalize $\mathcal{H}$, yielding an effective Hamiltonian $\mathcal{H}_\text{eff}$ that operates within the either one of the two subspaces while incorporating the effects of the states of the other through perturbative corrections.

The decoupling is achieved by a unitary transformation $U = e^{-S}$, where $S$ is an anti-Hermitian operator. This operator is carefully chosen to cancel the couplings between the low- and high-energy subspaces to a desired perturbative order. The transformed Hamiltonian is given by $\mathcal{H}_\text{eff} = U \mathcal{H} U^\dagger$. By expanding this expression using the Baker-Campbell-Hausdorff formula, a perturbative series for $\mathcal{H}_\text{eff}$ is derived in terms of commutators involving $S$
\begin{align}
\mathcal{H}_\text{eff} &= e^{-S} \mathcal{H} e^{S} \label{eq: time independent rotation}\\
 &= \mathcal{H} + [\mathcal{H}, S] + \frac{1}{2}[[\mathcal{H}, S], S] + \cdots \\
 & = \sum_{i=0}\left(\mathcal{H}^{(i)} +[\mathcal{H}^{(i)}, S] +  \frac{1}{2}[[\mathcal{H}^{(i)}, S], S] + \cdots \right)+ \sum_{j=1}\left( V^{(j)}  +  [V^{(j)}, S] + \cdots\right). \label{eq: H_eff full expansion}
\end{align}
To achieve block-diagonalization, $S$ is typically expanded as $S = \sum_j S^{(j)}$, with each term $S^{(j)}$ corresponding to a specific order of perturbation. At each order, $S^{(j)}$ is chosen to nullify the off-diagonal terms that couple $\ket{L}$ and $\ket{H}$ states, leading to an iterative construction of $\mathcal{H}_\text{eff}$. At this stage it is important to acknowledge that the choice of $S$ is not unique, and additional conditions are often imposed to ensure a well-defined transformation. Two commonly used conditions are (i) requiring $S$ to have a block-off-diagonal structure or (ii) minimizing the deviation of $U$ from the identity operator. While these conditions coincide for the standard SWT formalism presented in this section, variations arise in extensions. A particular case of this is multi-block diagonalization, where the identification of more than two substantial separations may indicate that a transformation capable of separating more than two blocks at the time may be more useful (see Sec.~\ref{sec: multi-block introduction}).

In standard SWTs, the effective Hamiltonian up to second order is derived by substituting the series expansion of $S$ into the perturbative expansion of $\mathcal{H}_\text{eff}$. For instance, at first order, $S^{(1)}$ satisfies the condition 
\begin{align}
    [\mathcal{H}^{(0)}, S^{(1)}] = -V^{(1)},  \label{eq: condition on regular SW}
\end{align}
while higher-order terms involve more complex commutators and interactions. Extending the SWT to time-dependent systems introduces additional challenges. This is especially true when the perturbative order of the rate of change of $S^{(j)}$ remains of order $j$ (i.e. $\frac{\partial S^{(j)}}{\partial t} \sim S^{(j)}$). In such cases, the condition imposed on the time-dependent generator $S^{(1)}(t)$ becomes the differential equation~\cite{SW_TD_subharmonic_transition_and_block_siegert}
\begin{align}
    [\mathcal{H}^{(0)}, S^{(1)}(t)] = -V^{(1)}(t)+i\hbar\frac{\partial S^{(1)}(t)}{\partial t}. \label{eq: condition on regular TDSW}
\end{align}

Equations~(\ref{eq: condition on regular SW}) and~(\ref{eq: condition on regular TDSW}) present the textbook example conditions required to perform regular SWTs up to second order. However, the SWT can in theory be declinated in a variety of different ``flavor''. This is often achieved by modifying the conditions imposed on each order of the generator $S$. For example, by requiring $\mathcal{H}^{(0)}$ to be free of degeneracies, it is possible to establish a set of conditions required for the full diagonalization of any perturbed Hamiltonian. In these cases, the condition imposed on $S$ for a second order time-independent transformation becomes
\begin{align}
    [\mathcal{H}^{(0)}, S^{(1)}] = -\mathcal{P}_\text{off}\left(\mathcal{H}^{(1)} + V^{(1)} \right), \label{eq: condition on full diagonalization SW}
\end{align}
where $\mathcal{P}_\text{off}(.)$ is defined as to project operators within their off-diagonal subspaces. Regardless of the chosen ``flavor'', the equations defining the conditions imposed on $S^{(j)}$ only vary on the couplings one wishes to eliminate up to the $j^\text{th}$ order. 

Historically, solutions for $S^{(j)}$ were often dependent on the dimensionality of the system and the complexity of $V^{(j)}$. In finite-dimensional systems, $S^{(j)}$ was typically determined by matching matrix elements, while infinite-dimensional cases required additional assumptions or truncations. Recently, a universal framework for constructing $S$ has been developed, as outlined in Ref.~\cite{universal_solution_paper}. This framework addresses limitations of earlier approaches, providing a systematic methodology for any perturbative Hamiltonian. The development of computational tools like SymPT leverages these advancements, automating the derivation of $S$ and enabling analysis of complex systems without requiring Hilbert space truncations.

\subsection{Block-diagonalization and the least action condition} \label{sec: multi-block introduction}
While SWT methods have been extensively developed, challenges remain for multi-block transformations. It was recently shown that the conditions of minimizing the action of $U$ on $\mathcal{H}$ and imposing a block-off-diagonal structure on $S$ do not always coincide~\cite{DiVicenzo}. While block-off-diagonal conditions are straightforward to handle using existing methods, minimizing the action on $\mathcal{H}$ requires additional formalism, which has, until now, not been fully developed. In this section we formalize the ideas behind block-diagonalization, elucidating the concept of imposing conditions on $S$ generator, whilst also presenting an answer to the questions left open in~\cite{DiVicenzo}. 

In general, an exact block diagonalization of a given Hamiltonian $\mathcal{H}$ is achieved via unitary transformations  
\begin{equation}
    \mathcal{H}_\text{block} = U \mathcal{H} U^\dagger,
\end{equation}
where the unitary operator $U$ is defined as $U = e^{-S}$ with $S$ being an anti-hermitian operator. As stated in the previous section, provided a fixed block-diagonal structure for $\mathcal{H}_\text{block}$, it is impossible to determine a unique solution for $S$. In standard SWTs this is often resolved by imposing $\mathcal{B}(S) =0$ where
\begin{equation}
    \mathcal{B}(S) \equiv  
    \begin{cases}  
    S_{ij}, & \text{if } i, j \in \text{same block}, \\  
    0, & \text{if } i, j \in \text{different blocks}.
    \end{cases} . 
\end{equation}
However a more physically understandable condition, first presented in~\cite{Cederbaum_1989}, is to impose that $U$ performs no operation on $\mathcal{H}$ other than block diagonalizing it. This condition is then formulated by conditioning the Euclidian norm between $U$ and the identity operator $I$ to be minimized
\begin{align}
    ||U - I|| =\mathrm{min}. \label{eq: LA condition}
\end{align}
Upon imposing this condition on $U$ (and by consequence imposing conditions on $S$), the resulting unitary operator is given by  
\begin{equation}
    U^\dagger = X^\dagger \mathcal{B}(X) \left\{\mathcal{B}(X^\dagger) \mathcal{B}(X)\right\}^{-\frac{1}{2}}, \label{eq: multi-block unitary}
\end{equation}
where $X$ is the operator that fully diagonalizes $\mathcal{H}$. Under the constraint of Eq.~\ref{eq: LA condition}, $U$ represents a full diagonalization followed by a ``back rotation'' to achieve block diagonalization of the Hamiltonian. While previous work, such as~\cite{DiVicenzo}, applied this approach for perturbative block diagonalization, only equations up to the third order are presented, leaving open the question of whether a general iterative approach to any order could be implemented. In Appendix~\ref{sec: multi-block generator}, we address this by presenting a generator capable of deriving closed-form expressions for $S^{(j)}$ to any desired order. This has been implemented in SymPT to facilitate these advanced transformations. This capability represents a significant step forward in the versatility of the SWT, enabling applications to a broader range of transformations.

\subsection{Algorithm for standard SWTs}\label{sec: algorithm SWT}

\begin{figure}[h]
    \centering
    \includegraphics[width=\linewidth]{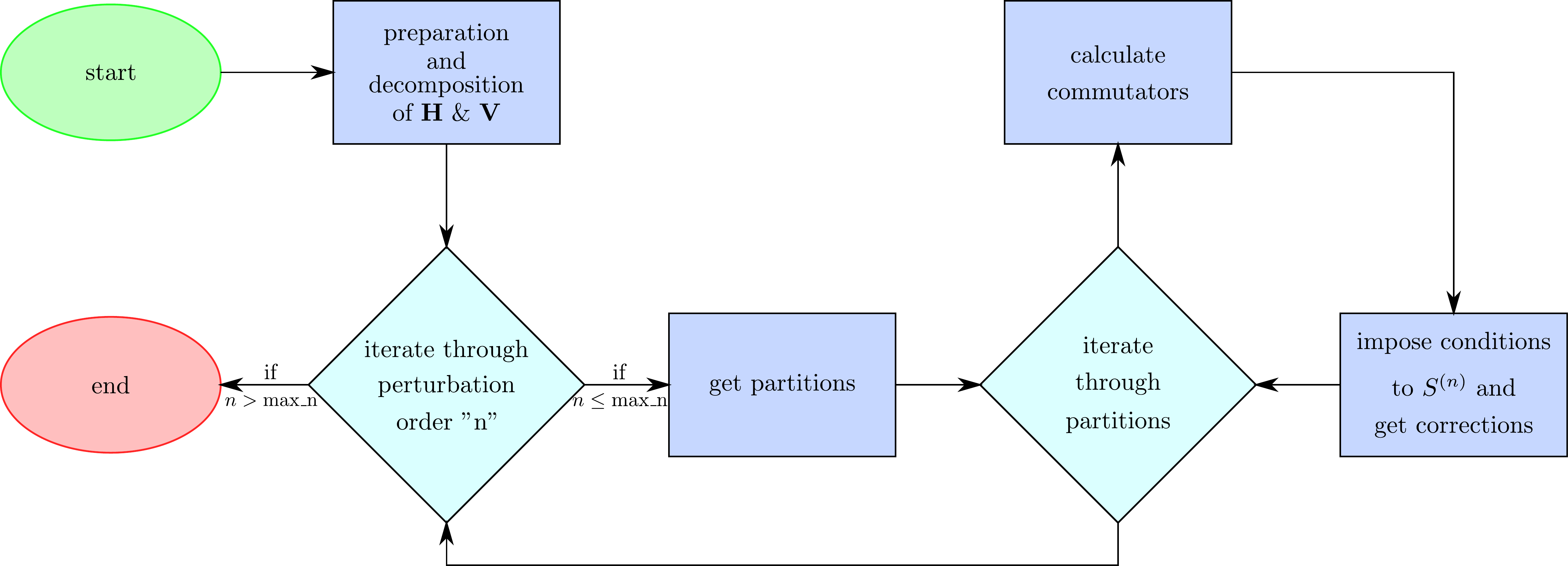}
    \caption{A schematic flux diagram of the algorithm implemented for the standard SWT as well as for the FD and ACE routine. These three routines are mostly equivalent, only differing at the step of imposing conditions to $S^{(n)}$ and obtaining the relevant corrections}
    \label{fig: flux diagram}
\end{figure}

SymPT utilizes the standard SWT routine to derive an effective Hamiltonian for quantum systems experiencing perturbative interactions. This algorithm (see Fig.~\ref{fig: flux diagram}) focuses on systematically block-diagonalizing the original Hamiltonian, as expressed in Eq.~(\ref{eq: decomposed hamiltonian}). The procedure adheres to the formalism outlined in Sec.~\ref{sec: the schrieffer-wolff transformation}, which necessitates that the user has predefined the block structure of the system. Consequently, the input Hamiltonian must be decomposed into two components: the unperturbed Hamiltonian $\mathbf{H} = \sum_{i=0}\mathcal{H}^{(i)}$ and the perturbation operator $\mathbf{V} = \sum_{j=1}V^{(j)}$. Note that this routine can implement both time dependent and time independent transformations. The presence of time dependent terms is automatically detected by SymPT, and thus no additional step is required by the user. 

The routine begins by preparing these input components, organizing them by their respective perturbative orders. This preparatory step allows the zeroth-order correction to the effective Hamiltonian to be directly stored (see Sec.~\ref{sec: How to use} for additional detail on how pertubative orders are handled in SymPT).
Subsequently, the algorithm generates a set of partitions for each perturbative order $n$. These partitions are ordered collections of integers summing to $n$, such as $(3)$, $(2,1)$, $(1,2)$, $(1,1,1)$ and $(0,3)$ for $n=3$. Each partition corresponds to specific nested commutators (obtained as exemplified in Eq.~(\ref{eq: H_eff full expansion})), with its length minus one representing the commutator's nestedness. The positional values within a partition denote the order of the operators involved. For instance, the partition $(1,1,1)$ could correspond to the commutators $[\mathcal{H}^{(1)}, S^{(1)}]^{(2)}$ or $[V^{(1)}, S^{(1)}]^{(2)}$, depending on the context. The partitions then serve two main purposes: firstly, these ensure comprehensive inclusion of all contributions at each order, secondly, the partitions can be used to index already computed commutators. With this, it is possible to easily cache in memory already computed nested commutators, thus drastically reducing the number of products to be computed. 

Iteratively, for each perturbative order, SymPT uses these partitions to determine the operator conditions on $S^{(n)}$, the anti-Hermitian generator of the transformation (as specified in Eq.~(\ref{eq: condition on regular SW})). Once these conditions are resolved, the solution for $S^{(n)}$ is computed using a method that leverages the theoretical results described in~\cite{universal_solution_paper}. These solutions are stored in memory and subsequently used to compute the $n$-th order correction to the effective Hamiltonian.

\subsection{Algorithm for FD and ACE} \label{sec: algorithm FD and ACE}

In addition to the standard SWT routine, SymPT implements routines for full diagonalization (FD) and arbitrary-coupling elimination (ACE) (for both time dependent and time independent systems). While the underlying algorithm shares similarities with the SWT process described above, key distinctions arise from the conditions imposed on the generator $S$ at each order, as discussed in Sec.~\ref{sec: the schrieffer-wolff transformation}. In particular note that both implementations enforce the condition of a block-off diagonal structure for $S$ (see Sec.~\ref{sec: algorithm LA} for additional details on performing block-diagonalization imposing least action conditions). 

The FD routine focuses on diagonalizing the Hamiltonian entirely, disregarding its block structure. This means users are not required to decompose the Hamiltonian into separate components and can instead provide the full Hamiltonian. The algorithm processes the input Hamiltonian by decomposing it into perturbative orders and computing the necessary commutators based on the generated partitions. The resulting terms are separated into diagonal and off-diagonal components. While diagonal components contribute directly to the effective Hamiltonian, off-diagonal terms define the conditions for $S^{(n)}$. Once these conditions are satisfied, $S^{(n)}$ is used to compute corrections to the effective Hamiltonian, ensuring complete diagonalization.

The ACE routine requires an additional \block object that specifies the couplings to be eliminated (see Sec.~\ref{sec: How to use} for additional details). This flexibility allows users to target specific off-diagonal elements for removal, provided these elements are sufficiently small compared to the coupled energies. As in the FD routine, partitions are generated for each order, and commutators are computed. The results are filtered based on the specified \block object, ensuring that only designated couplings contribute to the operator conditions for $S^{(n)}$. By iteratively applying these transformations, the ACE routine effectively eliminates arbitrary couplings while preserving the system's overall structure.

\subsection{Algorithm for LA multi-block transformations} \label{sec: algorithm LA}
In this section we present the algorithm implemented to perform multi-block diagonalization of provided Hamiltonians. Although the ACE routine could, in theory, be used to perform multi-block diagonalization, this would be performed by imposing a block-off diagonal structure onto the generator $S$. As mentioned in Sec.~\ref{sec: the schrieffer-wolff transformation} this is not the only possible condition that one could impose on $S$. It is thus necessary to provide the user the possibility of performing these class of transformations imposing a ``least-action'' (LA) condition instead. Note that a brief description of the mathematical formalism behind LA block-diagonalizations of Hamiltonians is presented in~Sec.~\ref{sec: multi-block introduction}.

As per the algorithms presented in Sec.~\ref{sec: algorithm FD and ACE}, to perform transformations using the LA method, the user is required to provide both the Hamiltonian and an additional \block object encoding the information of the off-diagonal blocks to be eliminated. The LA routine then starts by making use of the FD routine (see Sec.~\ref{sec: algorithm FD and ACE}) to obtain the anti-hermitian operators $Z^{(j)}$ (see Appendix~\ref{sec: multi-block generator}) required to generate the unitary $X$ (first introduced in Eq.~(\ref{eq: multi-block unitary})) that fully-diagonalizes system Hamiltonian. The routine then generates a new set of partitions based on Eq.~(\ref{eq: partitions set P}), which are used in combination with Eq.~(\ref{eq: multiblock generator}) to obtain the expressions for the generator of the unitary transformation $U$. Upon obtaining each order of the generator of $U$, the Hamiltonian is rotated using the  Baker-Campbell-Hausdorff expansion presented in Eq.~(\ref{eq: H_eff full expansion}).

\section{Examples}
\label{sec: Examples}

\subsection{EDSR in a slanting Zeeman field} \label{sec: rabi model example}
An example scenario where the SWT is useful is that of a spin qubit  in a slanting magnetic field coupled to an harmonic oscillator which is driven by a classical oscillating field \cite{EDSR1, EDSR2, EDSR3, EDSR4}. The Hamiltonian, in second quantization, is given by
\begin{equation}
\mathcal{H} =  \hbar \omega {{a}^\dagger} {a}  - \frac{\hbar \omega_{z}}{2}\sigma_{z} - \frac{\hbar \tilde{b}_{\mathrm{SL}}}{2} \left(a^\dagger + a\right) \sigma_x - \tilde{E}_0\cos\left(\omega_d t\right)\left(a^\dagger + a\right), \label{eq: EDSR HAmiltonian}
\end{equation}
where $\omega$ and $\omega_z$ are the resonator and qubit frequencies respectively, $\tilde{b}_{\mathrm{SL}}$ is proportional to the magnitude of the slanted magnetic field, while $\tilde{E}_0$ and  $\omega_d$ are instead proportional to the amplitude and frequency of the classical oscillating field. By applying a perturbative transformation to Eq.~(\ref{eq: EDSR HAmiltonian}) it is possible to derive an effective frame in which the oscillating field induces a change over time of the magnetic field experienced by the spin.
This is the reason for the name of the phenomena, electric dipole spin resonance (EDSR), which is a way to achieve spin control without the need for oscillating magnetic fields. In this example we consider the perturbative regime characterized by $\tilde{b}_\text{SL} \sim \tilde{E}_0 \ll |\omega \pm \omega_z|$. We use two approaches to obtain an effective qubit Hamiltonian: (i)  a time independent SWT of the undriven part of Eq.~(\ref{eq: EDSR HAmiltonian}) followed by a rotation of the driving term into the transformed frame; (ii) a time-dependent SWT of the total Hamiltonian. 

(i) Using SymPT's standard SWT routine (see Sec.~\ref{sec: algorithm SWT})  we obtain the undriven transformed Hamiltonian up to second order  
\begin{equation}
\mathcal{H}^{(2)}_\text{eff} = \frac{\hbar \omega_{z} \tilde{b}_{\mathrm{SL}}^{2}}{4 \left(\omega^{2} - \omega_{z}^{2}\right)} \left({{a}^\dagger}^2 + a^2 + 2a^\dagger a + 1\right)\sigma_z.
\end{equation}
To include the effects of the oscillating classical field, we make use of SymPT in-build functionalities (see Sec.~\ref{sec: Effective frame}) to  rotate the driving term into the newly defined frame, which reads
\begin{equation}
 \mathcal{H}_\text{Drive} = - \tilde{E}_0 \cos{\left(\omega_{d} t \right)} \left({{a}^\dagger} + {a}\right) - \frac{\omega \tilde{E}_0 \tilde{b}_{\mathrm{SL}} }{\omega^{2} - \omega_{z}^{2}} \cos{\left(\omega_{d} t \right)} \sigma_x.
\end{equation}
An effective qubit Hamiltonian is then obtained by projecting the resulting total Hamiltonian onto the ground state of the harmonic oscillator
\begin{equation}
\begin{aligned}
    \mathcal{H}_\text{qubit} =& - \frac{\hbar\omega_\text{qubit}}{2}\sigma_z - \frac{\omega \tilde{E}_0 \tilde{b}_{\mathrm{SL}} }{\omega^{2} - \omega_{z}^{2}} \cos{\left(\omega_{d} t \right)} \sigma_x, \label{eq: EDSR qubit ham}
\end{aligned}
\end{equation}  
where
\begin{equation}\label{eq:EDSRomega_qubit}
\omega_\text{qubit} \equiv \omega_z \left(1 - \delta_z\right), \quad \delta_z \equiv \frac{\tilde{b}_{\mathrm{SL}}^{2}}{2 \left(\omega^{2} - \omega_{z}^{2}\right)}.
\end{equation}

(ii) Using a time-dependent SWT (see Sec.~\ref{sec: the schrieffer-wolff transformation}) of the total Hamiltonian presented in Eq.~(\ref{eq: EDSR HAmiltonian}), the second order effective Hamiltonian, in the limit of $\tilde{b}_\text{SL} \sim \tilde{E}_0 \ll |\omega \pm \omega_z|$ and  $\tilde{b}_\text{SL} \sim \tilde{E}_0 \ll |\omega \pm \omega_d |$, takes the form  
\begin{align}
\mathcal{H}^{(2)}_\text{eff} =& \frac{\hbar \omega_{z} \tilde{b}_{\mathrm{SL}}^{2}}{4 \left(\omega^{2} - \omega_{z}^{2}\right)} \left({{a}^\dagger}^2 + a^2 + 2a^\dagger a + 1\right)\sigma_z \\
&- \frac{\omega \tilde{E}_0 \tilde{b}_{\mathrm{SL}}}{2} \left(\frac{1}{\omega^2-\omega_z^2} + \frac{1}{\omega^2-\omega_d^2}\right) \cos{\left(\omega_{d} t \right)} \sigma_x.
\end{align}
Projecting onto the harmonic oscillator's ground state yields
\begin{equation}
\begin{aligned}
    \mathcal{H}_\text{qubit} =& - \frac{\hbar\omega_\text{qubit}}{2}\sigma_z - \frac{\omega \tilde{E}_0 \tilde{b}_{\mathrm{SL}} }{2} \left(\frac{1}{\omega^2-\omega_z^2} + \frac{1}{\omega^2-\omega_d^2}\right) \cos{\left(\omega_{d} t \right)} \sigma_x. \label{eq: EDSR qubit ham TD}
\end{aligned}
\end{equation}

Comparing Eq.~(\ref{eq: EDSR qubit ham}) with Eq.~(\ref{eq: EDSR qubit ham TD}) we note that, although the qubit frequency $\omega_\text{qubit}$ remains unvaried, the amplitude of the oscillating field in Eq.~(\ref{eq: EDSR qubit ham TD}) changes when compared to Eq.~(\ref{eq: EDSR qubit ham}). This difference arises because of the time-dependence of the frame defined by the second approach. In this case the time dependent transformation partly accounts for the rotation of the drive, resulting in a distinct effective drive in the TLS. Nevertheless, when the drive is resonant with the effective qubit frequency, $\omega_d = \omega_\text{qubit}$, it is possible to perform a series expansion around $\delta_z = 0$ and show that both approaches yield the same effective Hamiltonian up to order $\mathcal{O}(\tilde{b}_\text{SL}^2)$.

\subsection{Transmon coupled to resonator} \label{sec: transmon qubit example}
In this section we present the results obtained using SymPT to analyse a system comprised by a transmon system coupled to a superconducting resonator
\begin{align}
    \mathcal{H} = \omega_t a_t^\dagger a_t  +\omega_ra_r^\dagger a_r  + \frac{\alpha}{2}a_t^\dagger a_t^\dagger a_ta_t - g\left(a_t^\dagger - a_t \right)\left(a_r^\dagger - a_r\right) \label{eq: hamiltonian transmon resonator},
\end{align}
where $\left[a_r,a_r^\dagger\right] = \left[a_t,a_t^\dagger\right]=1$ and where the respective number operators are given by $N_r = a_r^\dagger a_r$ and $N_t = a_t^\dagger a_t$. Similar systems have been extensively studied in the literature~\cite{Benjamins_paper, Transmon_resonator_1, Transmon_resonator_2, transmon_resonator_3} in the dispersive regime $|g|\ll |\omega_r-\omega_t-n_t\frac{\alpha}{2}|$ $\forall n_t\in \mathbb{Z}^\geq$, however, due to their complexity, perturbative calculations of these systems often require either a truncation of the bosonic subspaces, or otherwise extensive and complicated computations.  
Dispersive analysis of these systems using standard SWTs 
achieves the separation of even-odd numbered resonator excitation subspaces. 
Fully separating all the subspaces corresponding to a given resonator excitation number from each other,
requires a different implementation of the SWT.
The use of the ACE routine (see Sec.~\ref{sec: algorithm FD and ACE}) still results in the appearance of two photon process terms (i.e. proportional to $a_r^2$), which are at times undesired. In this case, SymPT can be employed to perform a FD of the system (see Sec.~\ref{sec: algorithm FD and ACE}) to determine a suitable frame in which all those subspaces are fully separated and the two photon processes are suppressed. Here we find the second order correction (note that no first order correction exists for this system) to be
\begin{align}
    \mathcal{H}_\text{eff}^{(2)} &= \Omega_ta^\dagger_ta_t +\Omega_r a^\dagger_ra_r + \alpha'\left(a_t^\dagger a_t\right)^2  + \mathcal{H}^{(2)}_{LCK} +\mathcal{H}^{(2)}_{NLCK},\label{eq: second order correction to transmon resonator}
\end{align}
where $\Omega_t$, $\Omega_r$ are second order corrections to the transmon, resonator frequencies respectively and $\alpha'$ is the correction to the transmon's anharmonicity (see Appendix~\ref{sec: transmon resonator corretions} for the expanded form of these terms). In Eq.~(\ref{eq: second order correction to transmon resonator}) we also note the appearance of both linear and non-linear cross-Kerr interaction terms~\cite{Benjamins_paper}
\begin{align}
    \nonumber \mathcal{H}^{(2)}_{LCK}=&g^2 \left[\frac{2}{N_{t} \alpha - \alpha + \omega_{r} + \omega_{t}} + \frac{2 }{N_{t} \alpha - \alpha - \omega_{r} + \omega_{t}} - \frac{2 }{N_{t} \alpha + \omega_{r} + \omega_{t}} + \right .\\
    \nonumber & \hspace{0.5 cm}- \frac{2 }{N_{t} \alpha - \omega_{r} + \omega_{t}} + \frac{\alpha  -  \omega_{r} -  \omega_{t}}{\left(N_{t} \alpha - \alpha + \omega_{r} + \omega_{t}\right)^{2}} + \frac{\alpha  +  \omega_{r} -  \omega_{t}}{\left(N_{t} \alpha - \alpha - \omega_{r} + \omega_{t}\right)^{2}} +\\
    & \left . \hspace{0.5 cm}+ \frac{\alpha +  \omega_{r} +  \omega_{t}}{\left(N_{t} \alpha + \omega_{r} + \omega_{t}\right)^{2}} + \frac{\alpha  - \omega_{r} +  \omega_{t}}{\left(N_{t} \alpha - \omega_{r} + \omega_{t}\right)^{2}}\right]N_tN_r,\\
    \nonumber \mathcal{H}^{(2)}_{NLCK}=&\alpha g^{2} \left[- \frac{1}{\left(N_{t} \alpha - \alpha + \omega_{r} - \omega_{t}\right)^{2}} - \frac{1}{\left(N_{t} \alpha - \alpha - \omega_{r} - \omega_{t}\right)^{2}} + \right.\\
    & \hspace{0.9 cm} \left. +\frac{1}{\left(N_{t} \alpha + \omega_{r} - \omega_{t}\right)^{2}} + \frac{1}{\left(N_{t} \alpha - \omega_{r} - \omega_{t}\right)^{2}}\right]
N_t^2N_r 
\end{align}
These terms, encode the effective interplay between the transmon and resonator subsystems at second order. Specifically, the linear cross-Kerr term introduces a dependence of the resonators photon number on the effective transmon's energy levels. Similarly, the non-linear cross-Kerr term introduces higher-order dependencies on the resonator's photon number in the non-linear terms of the transmon system. These terms are often leveraged for the study of controlled interactions between transmons and microwave resonators, becoming especially relevant in the analysis of dispersive measurements and decoherence mechanisms of these systems.

As a final remark, it is important to highlight that while the specific choice of SWT flavor may appear inconsequential, since second-order solutions often yield relatively similar results, the newly defined frame is not. Rotating operators into different frames can thus lead to distinct outcomes already at second order. To address this, SymPT retains detailed information about the frame associated with the selected transformation. This enables the rotation of any desired operator into the newly defined frame, as demonstrated in Sec.~\ref{sec: rabi model example}.

\subsection{Block-diagonalizations of stochastic Hamiltonians} \label{sec: BD random matrix example}
\begin{figure}[H]
\centering
\minipage{0.3\textwidth}
  \includegraphics[width=\linewidth]{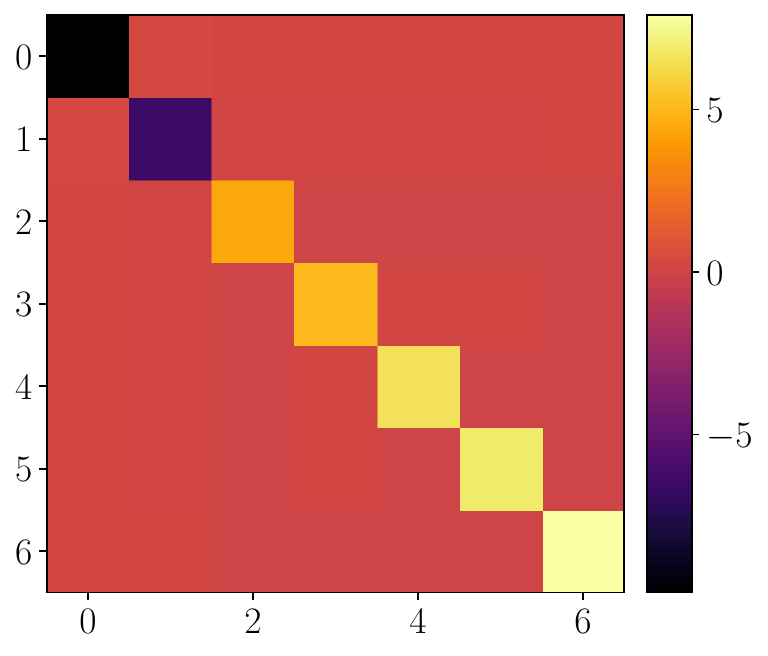}
  \label{fig: LA_non_masked}
\endminipage
\hspace{0.3 cm}
\minipage{0.3\textwidth}
  \includegraphics[width=\linewidth]{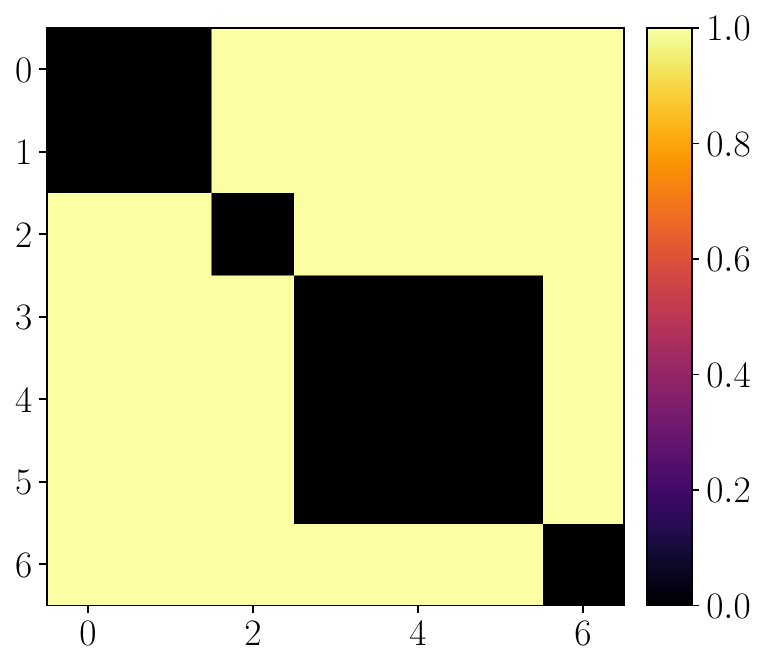}
  \label{fig: LA mask}
\endminipage
\hspace{0.3 cm}
\minipage{0.3\textwidth}%
  \includegraphics[width=\linewidth]{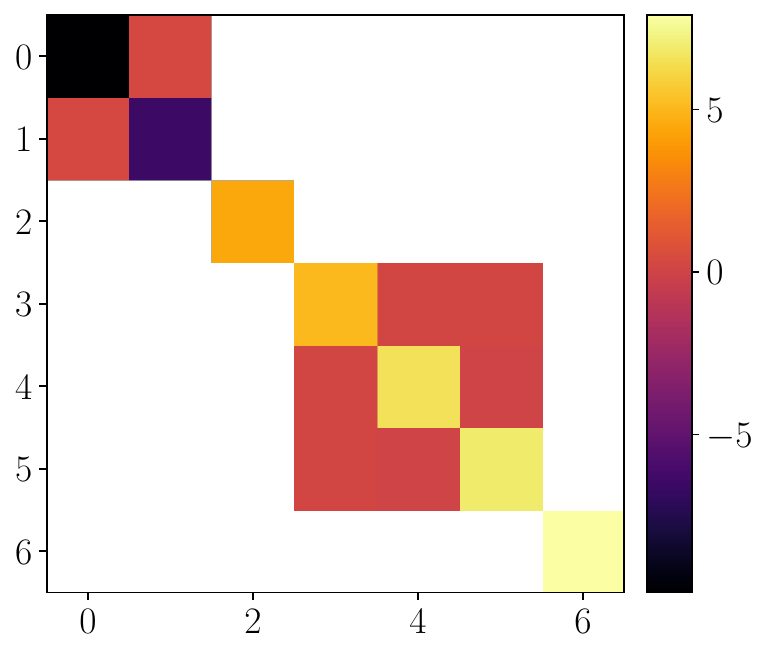}
  \label{LA masked}
\endminipage
\caption{\textbf{Panel 1}: The stochastic Hamiltonian before the transformation. This was created as to contain randomly generated second order perturbative elements everywhere outside of a (also randomly generated) block structure. \textbf{Panel 2}: The mask used to determine the block structure of the transformed Hamiltonian. This mask was generated by randomly selecting the number of blocks and their dimensionality. \textbf{Panel 3}: The transformed Hamiltonian up to eigth order. The matrix elements with value zero were set to have white color to distinguish them from otherwise low-valued elements.}
\label{fig: BD Random Hamiltonian}
\end{figure}
In several interesting scenarios~\cite{block_diagonal_example_1, block_diagonal_example_2, block_diagonal_example_3}, the energy spectrum of the system may exhibit multiple substantial separations rather than a single, large energy gap, leading to the identification of multiple distinct "block subspaces". In these cases, the standard SWT routine discussed in the previous sections cannot be used to adequately capture the effective behavior of the system studied. Instead, a modified approach, capable of accounting for the unique structure of each of these cases, must be formulated. As introduced in Sec.~\ref{sec: multi-block introduction}, one way to address this is by imposing specific conditions on the anti-Hermitian generator $S$ to derive a unique transformation that block-diagonalizes the Hamiltonian. In this section, we focus on applying this methodology to the block-diagonalization of stochastic matrices, utilizing the LA condition defined earlier. Although the results presented here are centered on finite systems, this approach can also be extended to systems containing bosonic subspaces without necessitating Hilbert space truncation.
\begin{figure}[H]
\centering
  \includegraphics[width=0.7\linewidth]{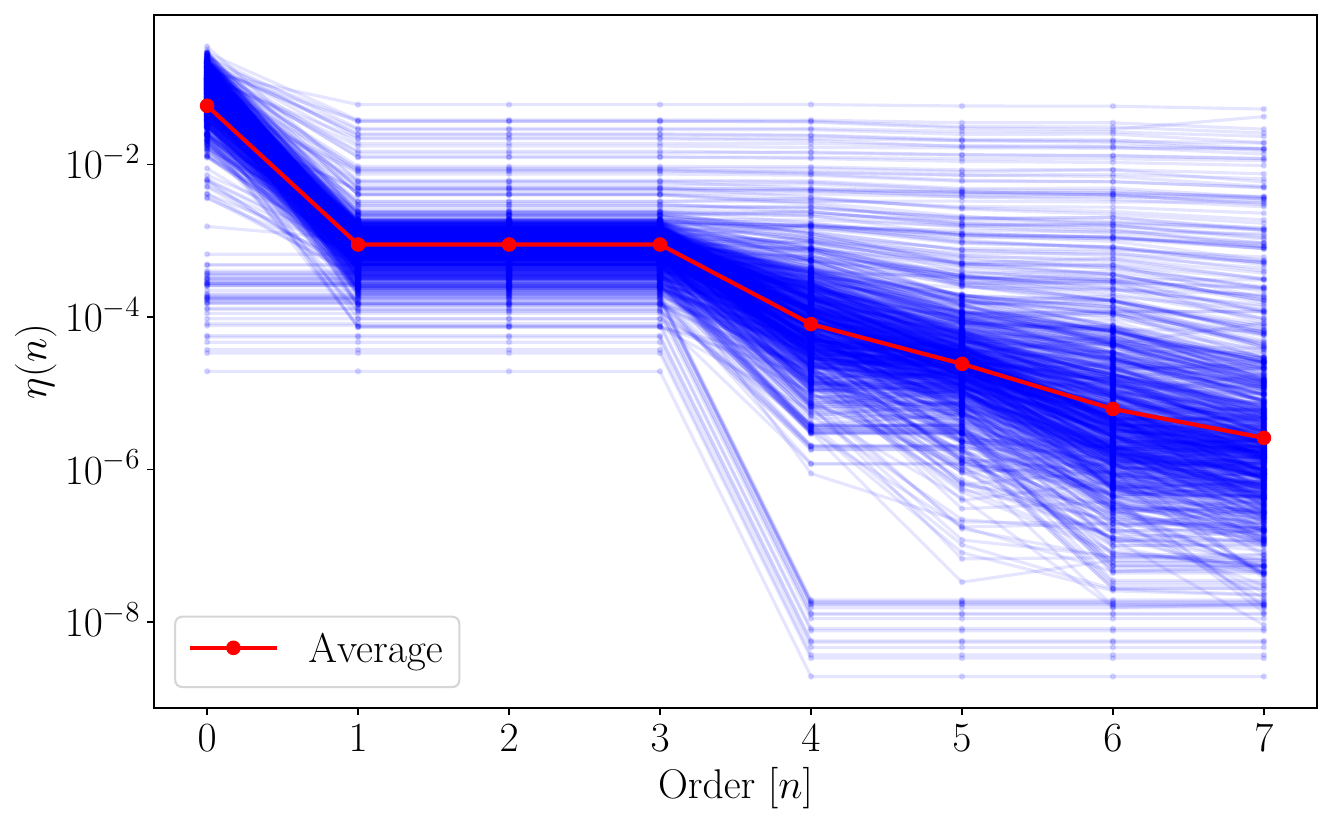}
\caption{The relative spectral distance $\eta(n)$ as a function of the transformation order $n$, shown for 900 stochastic Hamiltonians. The plot compares the transformed Hamiltonians $\mathcal{H}^{(n)}_\text{LA}$ obtained using the LA condition with the exact block-diagonalized Hamiltonians $\mathcal{H}_\text{exact}$, computed numerically. For each Hamiltonian, the dimensionality, block structure, and the values of the system parameters were randomly generated within specified ranges to ensure diverse configurations.}
\label{fig: LA relative distance}
\end{figure}
An example Hamiltonian analyzed in this section is displayed in panel 1 of Fig.\ref{fig: BD Random Hamiltonian}. Panels 2 and 3 illustrate the mask applied to the system and the resulting transformed Hamiltonian, respectively. This example demonstrates SymPT's capability to implement this class of transformations under the LA conditions introduced in Sec.\ref{sec: multi-block introduction}. Further evidence is provided in Fig.\ref{fig: LA relative distance}, which depicts the relative spectral distance $\eta(n)=\frac{||\mathcal{H}_\text{exact} - \mathcal{H}^{(n)}_\text{LA}||}{||\mathcal{H}_\text{exact}||}$, where $||\cdot||$ represents the spectral norm. This metric compares the accuracy of the SymPT routine across different orders $n$ for 900 stochastic Hamiltonians with their exact counterparts, derived via numerical evaluation of the unitary transformation defined in Eq.(\ref{eq: multi-block unitary}). To achieve a clear energy separation between multiple subspaces of the generated Hamiltonians, the coupling terms within the diagonal blocks were defined as first-order interactions, while those outside the blocks were set to second order. This distinction explains the plateau observed between $n=1$ and $n=3$, as no corrections are expected for these orders due to the hierarchy of the coupling terms.
\subsection{ACE of stochastic Hamiltonian} \label{sec: random matrix example}
\begin{figure}[H]
\centering
\minipage{0.31\textwidth}
  \includegraphics[width=\linewidth]{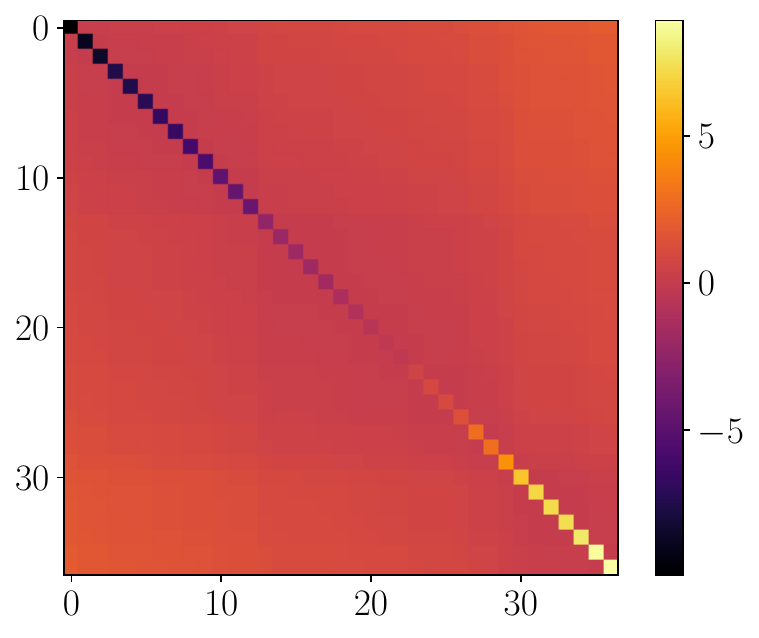}
  \label{fig: spectrum SymPT non-mask}
\endminipage
\hspace{0.3 cm}
\minipage{0.31\textwidth}
  \includegraphics[width=\linewidth]{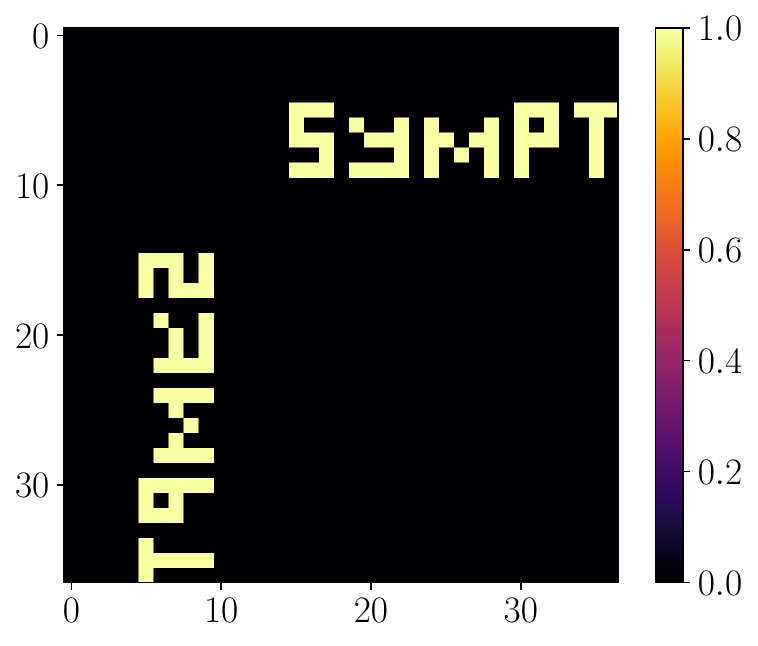}
  \label{fig: Ace mask}
\endminipage
\hspace{0.3 cm}
\minipage{0.31\textwidth}%
  \includegraphics[width=\linewidth]{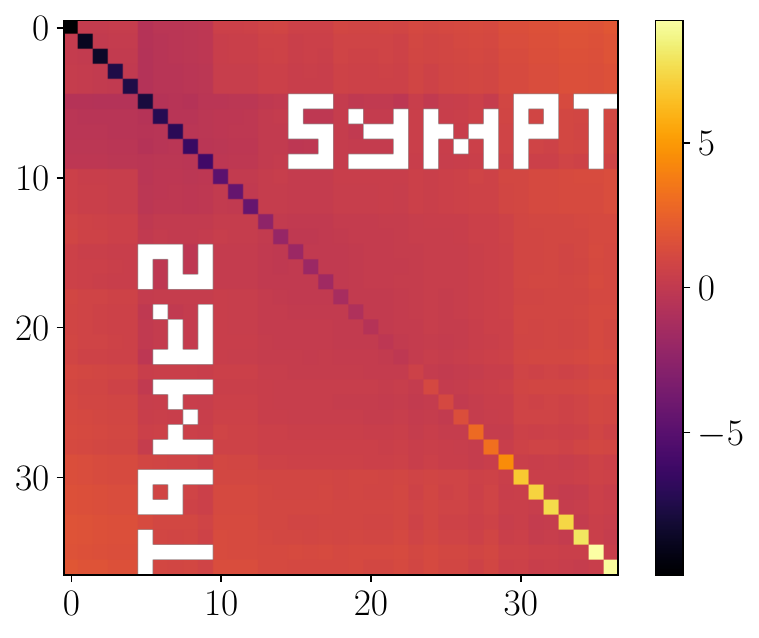}
  \label{Filtered Outcome}
\endminipage
\caption{\textbf{Panel 1}: The stochastic Hamiltonian before the transformation. This was created by arranging the randomly generated diagonal elements in increasing order, and as to contain first order perturbative elements everywhere outside of the main diagonal. \textbf{Panel 2}: The implemented mask targetting the off-diagonal couplings to eliminate. \textbf{Panel 3}: The transformed Hamiltonian up to third order. The matrix elements with value zero were set to have white color to distinguish them from otherwise low-valued elements.}
\label{fig: Random Hamiltonian}
\end{figure}

In this section, we demonstrate the flexibility of SymPT by applying a third-order ACE transformation to a randomly generated Hamiltonian. This method showcases the capacity of SymPT to selectively target specific couplings for elimination or retention, allowing for tailored manipulations of Hamiltonian structures. The transformations leverage conditions imposed on the generator $S$, similar to the approach used for the full-diagonalization flavor discussed in Sec.~\ref{sec: the schrieffer-wolff transformation}. To perform arbitrary coupling selection, SymPT users define a mask that specifies the elements of the Hamiltonian to be eliminated during the transformation process (see Sec.~\ref{sec: algorithm FD and ACE}). This capability extends to systems with bosonic subspaces, circumventing the need for Hilbert space truncation, which is particularly advantageous for systems with infinite-dimensional subspaces.

Figure~\ref{fig: Random Hamiltonian} provides a comprehensive visualization of the process: in panel 1 we show the initial stochastic Hamiltonian before transformation. The matrix was constructed by arranging randomly generated diagonal elements in ascending order and populating all off-diagonal entries with first-order perturbative elements. On the other hand panel 2 of Fig.~\ref{fig: Random Hamiltonian} illustrates the mask applied to the Hamiltonian, targeting specific off-diagonal couplings for elimination. Lastly, panel 3 presents the transformed Hamiltonian up to third order. Here, matrix elements with a value of zero are displayed in white to distinguish them from non-zero low magnitude elements.

The results highlight the efficacy of SymPT in implementing ACE transformations under user-defined conditions. By providing fine-grained control over the couplings to be preserved or eliminated, SymPT facilitates tailored analyses of complex systems. The elimination of selected couplings aligns with perturbative goals, ensuring the resulting Hamiltonian retains the desired structure while adhering to the constraints imposed by the chosen order of transformation.

\newpage
\section{How to use}
\label{sec: How to use}

\subsection{The objects and their attributes} \label{sec: the objects}
This section provides an overview of the primary classes implemented in SymPT that users must be familiar with to fully utilize this tool. These classes and their methods are introduced in the order they are typically applied, following the workflow we recommend.
\subsubsection{RDSymbol}
We begin with the \rdsymbol  class. This class is designed to streamline the setup of the system to be transformed, by allowing user to define scalar, commutative quantities. Being a child class of the \texttt{sympy.symbol} object class, \rdsymbol inherits all the attributes of its parent, thus allowing specification of properties such as \texttt{real} and \texttt{positive}. A notable difference between \rdsymbol and its parent class is the introduction of the \texttt{order} attribute: the introduction of this attribute allows to establish how each initialized symbol scales with the system's perturbative terms. The \texttt{order} can assume any real value, though it is important to ensure that no negative or non-integer orders appear in the finalized Hamiltonian. As a case study example, to setup the system studied in Sec.~\ref{sec: rabi model example}, we initialize five instances of \rdsymbol:

\begin{lstlisting}[language=Python]
# ---------------- Defining the symbols ------------------
# Order 0
omegaz = RDSymbol('omega_z', real=True, positive=True)
omega = RDSymbol('omega', real=True, positive=True)
omegad = RDSymbol('omega_d', real=True, positive=True)

# Order 1
E0 = RDSymbol(r'\tilde{E}_{\mathrm{0}}', real=True, order=1)
bsl = RDSymbol(r'\tilde{b}_{\mathrm{SL}}', real=True, order=1)     
\end{lstlisting}
Here, only the terms $\tilde{b}_\text{SL}$ and $\tilde{E}_0$, are defined as perturbative (see Sec.~\ref{sec: rabi model example}). As a remark, note that for time dependent transformations, SymPT makes use of inbuilt definitions for the variables $\hbar$ and $t$ (i.e. the time variable). It is therefore a ''best-practice'' not to redefine these two variables, but instead directly import them from SymPT.
\subsubsection{RDBasis} \label{sec: RDbasis}
As per \rdsymbol, the \rdbasis class is implemented to aid the set up of otherwise complicated system's Hamiltonians. This class is designed to encode all necessary information about the finite Hilbert subspaces comprising the total system. While not strictly essential for SymPT’s functionality, \rdbasis provides a useful structure for organizing the system's Hamiltonian. Initialization of this class requires a unique \texttt{name} to distinguish it from other subspaces, as well as a \texttt{dim} attribute, which specifies the dimensionality of the subspace. Upon initialization, \rdbasis generates the set of generalized Gell-Mann matrices~\cite{Gell-Mann} required to span the operator space of the given dimensionality. For two-dimensional systems, these correspond to Pauli matrices, although for higher-dimensional spaces, formulating a Hamiltonian using these operators can be sometimes challenging. To assist with this, \rdbasis contains the \texttt{project()} method, which allows users to decompose \texttt{sympy.Matrix} objects into the basis operators stored within the initialized \rdbasis. 
As an example of the initialization of this class, consider once again the example presented in Sec.~\ref{sec: rabi model example}; there we defined a two-dimensional spin subspace as well as the bosonic creation and annhilation operators with:
\begin{lstlisting}[language=Python]
# ----------------- Defining the basis -------------------
# Spin basis: Finite 2x2 Hilbert space
Spin = RDBasis('sigma', 2)
s0, sx, sy, sz = Spin.basis # Pauli operators
# Boson basis: Infinite bosonic Hilbert space
a = BosonOp('a')
ad = Dagger(a)
\end{lstlisting} 
Note that the \texttt{s0,sx,sy,sz} objects initialized in the above examples are instances of the custom made \texttt{RDOperator} class. In short, this class is a child of the \texttt{sympy.quantum.Operator} class, and it therefore inherits all of its properties. This enables \texttt{RDOperator} instances to include an additional \texttt{.matrix} property, thus allowing for the study of finite system's operators in their matrix representation. However, this functionality is often not required for most of the functionalities included in SymPT. 
\subsubsection{EffectiveFrame} \label{sec: Effective frame}
Another important class to introduce is the \effectiveframe class. In general, this object is initialized after defining the system Hamiltonian with its perturbations and it is designed to setup the desired perturbative transformation. The initialization of this class depends on the ''flavor'' of the perturbative transformation desired. To setup an \effectiveframe aimed at performing a standard SWT, the Hamiltonian must be manually decomposed in the form of Eq.~(\ref{eq: decomposed hamiltonian}) (see Sec.~\ref{sec: algorithm SWT} for additional details). With this, the \effectiveframe can be initialized providing the block-diagonal Hamiltonian \texttt{H}, the perturbative couplings \texttt{V} between the blocks of \texttt{H}, and a list of \rdbasis objects representing the finite subspaces within the Hamiltonian. The \texttt{H} and \texttt{V} objects used to initialize \effectiveframe may either be \texttt{sympy.Matrix} instances or their projections onto the system’s finite subspace. 
To initialize an \effectiveframe for any other perturbative transformation, the initial decomposition can be avoided, and the total Hamiltonian can be fed into the \texttt{H} attribute of the \effectiveframe.

Once initialized, the transformation is executed with the \texttt{.solve()} method. By specifying the perturbation order and the transformation “flavor”, users can retrieve the transformed Hamiltonian via the \texttt{.get\_H()} method, selecting the output format without needing to recompute the solution each time. Referring to the example presented in Sec.~\ref{sec: rabi model example}, the effective frame and the final expression for the effective Hamiltonian are obtained with:
\begin{lstlisting}[language=Python]
# -------------- Defining the Hamiltonian ----------------
# Unperturbed Hamiltonian H0
H0 = hbar * omega0 * (ad*a + sp.Rational(1,2)) -sp.Rational(1,2) * omegaz * sz
# Perturbation Hamiltonians
V = - sp.Rational(1,2) * bsl * (ad + a) * sx
HE = - E0 * sp.sin(omega * t) * (ad + a)

# --------------  Deffining Effective Frame  ----------------
Eff_Frame = EffectiveFrame(H = H0, V = V + HE, subspaces=[Spin])
# SWT up to the second order
Eff_Frame.solve(max_order=2, method="SW")
# Obtaining the result in operator form
H_eff = Eff_Frame.get_H(return_form='operator')
\end{lstlisting}

As discussed in Secs.~\ref{sec: transmon qubit example} and~\ref{sec: random matrix example}, SymPT can perform perturbative transformations extending beyond the conventional SWT. These expanded functionalities are accessible via the \texttt{.solve()} method of the \texttt{EffectiveFrame} class. This allows for either a full diagonalization invoking \texttt{.solve(method="FD")}, the selective elimination of specific couplings with \texttt{.solve(method="ACE", mask = my\_mask)} as well as block diagonalizations imposing least action conditions with \texttt{.solve(method="LA", mask = my\_mask)} (see Sec.~\ref{sec: block object} for additional details on the \texttt{mask} variable). 

Additionally, \effectiveframe grants the user the access to additional functionalities. For instance, it is also possible to separate the obtained result into the obtained corrections for each different order. Once the effective Hamiltonian is computed, the corrections to each order can be obtained via the \texttt{.corrections()} method implementd within \effectiveframe. This returns a python dictionary object, whose keys indicate the correction order, and whose items are the respective corrections. Lastly, it is also possible to use \effectiveframe to rotate any given operator to the newly obtained frame. This is achieved via the \texttt{.rotate()} method. This function only requires the operator that the user wishes to rotate: this can be provided both as a \texttt{sympy.Matrix} instance or as an expression obtained via the \protect method discussed in Sec.~\ref{sec: RDbasis}. In this regard, note that it is also possible to rotate any operator up to any other order below the order of the obtained effective Hamiltonian.

\subsubsection{Block} \label{sec: block object}
In order to implement the ACE and LA routines, SymPT introduces the \block class. This object enables precise control over the couplings to be targeted by the transformation. Initializing a \block instance requires two inputs: \texttt{fin} and \texttt{inf}. The \texttt{fin} input specifies the elements within the finite subspace of the system to be eliminated. This can be provided both as a \texttt{sympy.Matrix} object comprised of $0$ and non-$0$ elements, or as an operator expression obatined via the \texttt{RDOperator} class (see Sec.~\ref{sec: RDbasis}). Whenever \texttt{fin} was to be input as such, it is required that user provides an ordered list of the subspaces inlcuded in the transformation. On the other hand, the \texttt{inf} parameter identifies the terms within the bosonic subspace to be addressed. These \block instances can be combined by summing them to form a “mask” expression, which is then supplied to the \texttt{.solve()} method, indicating the terms to be targeted in the transformation. This approach grants flexibility in targetting arbitrary couplings within the Hamiltonian that one wishes to eliminate.

Consider the transmon-resonator system discussed in Sec.~\ref{sec: transmon qubit example}. In many cases, studies of such systems do not require a full diagonalization but instead aim for a complete separation of all the subspaces corresponding to a given resonator excitation number. This separation can be achieved by providing a suitable mask to the ACE routine in SymPT. The following code demonstrates how to perform this separation up to second order:
\begin{lstlisting}[language=Python]
# ---------------- Defining the symbols ------------------
# Order 0
omega_t = RDSymbol('omega_t', order=0, positive=True, real=True)
omega_r = RDSymbol('omega_r', order=0, positive=True, real=True)
alpha   = RDSymbol('alpha', order=0, positive=True, real=True)
# Order 1
g = RDSymbol('g', order=1, positive=True, real=True)

# ----------------- Defining the basis -------------------
# Boson basis transmon: Infinite bosonic Hilbert space
a_t  = BosonOp('a_t')
ad_t = Dagger(a_t)
# Boson basis resonator: Infinite bosonic Hilbert space
a_r  = BosonOp('a_r')
ad_r = Dagger(a_r)

# -------------- Defining the Hamiltonian ----------------
H0 = omega_t * ad_t * a_t + omega_r * ad_r * a_r + sp.Rational(1,2) * alpha * ad_t * ad_t * a_t * a_t # Unperturbed Hamiltonian H0
V = -g * (ad_t - a_t) * (ad_r - a_r) # Interaction Hamiltonian V
# -------------- Defining the EffectiveFrame --------------
Eff_frame = EffectiveFrame(H0, V)
# -------------- Deffining the mask -----------------------
mask = Block(inf=a_r*a_t) +  Block(inf=ad_r*a_t) + Block(inf=a_t**2*a_r**2)+ Block(inf=ad_t**2*a_r**2)

# -------------- Calculate the effective model ------------
Eff_frame.solve(max_order=2, method="ACE", mask=mask)
H_eff_Mask = Eff_frame.get_H(return_form='operator')
\end{lstlisting}

\subsubsection{Other useful tools}
Together with the custom defined objects discussed in this section, SymPT also includes a variety of different tools for the analysis of the studied results. For instance, it is sometimes convenient to require the \texttt{.solve()} method to return a solution in which the finite components of the Hamiltonian expression are separated according to the bosonic operators multiplying them. This can be achieved by using \texttt{return\_form = "dict\_operator"} or \texttt{return\_form = "dict\_matrix"} (depending on whether the user would like their finite system to be represented in operator or matrix form). In these scenarios, the output takes the form of a python dictionary, where the "keys" are the bosonic operators multiplying the respective finite operator "items". To ease with the readability of these output forms, SymPT provides the user with the \texttt{display\_dict()} function. This is a tailored made function to enable a more readable printing of python dictionary objects.

Additionally, SymPT also provides the user with the \texttt{group\_by\_operators()} function. When applied to an expression in "operator form", \texttt{group\_by\_operators()} returns a dictionary separating the operators contained in the expression from their rescaling scalar factors. Such dictionary can then be easily printed via the use of the previously mentioned \texttt{display\_dict()} function.

Lastly, to ease the user in the creation of block off-diagonal masks, SymPT provides the user with the function \texttt{get\_block\_mask()}. This function requires an ordered list of integer numbers that are used to encode the dimensionality of each diagonal block in the Hamiltonian. 
\section{Conclusion}
In this work, we have presented SymPT, an analytical software tool designed to automate the SWT and its extensions for both time-independent and time-dependent systems. Additionally SymPT enables the systematic derivation of transformation generators and effective Hamiltonians at both operator and matrix levels. Built on a unified framework for the SWT~\cite{universal_solution_paper}, the tool supports a range of functionalities, including arbitrary coupling elimination, full diagonalization, and block diagonalization guided by least-action conditions, thus addressing an existing gap in the computation of effective Hamiltonians. These capabilities significantly expand the applicability of SymPT, making it a versatile solution for a broad class of quantum systems and transformations.

Despite these advancements, certain limitations remain. The performance of SymPT in extremely high-dimensional finite Hilbert spaces could benefit from further optimization, particularly through the implementation of parallelization or multithreading routines. The software’s current structure lends itself well to such enhancements, making this a promising avenue for future improvements. A more critical limitation lies in the absence of optimization strategies for reducing the number of commutators required during calculations. This aspect has been addressed in prior work~\cite{pymablock}, where efficient algorithms significantly decreasing computational overhead have been derived. Integrating similar optimizations into SymPT would enhance its scalability and computational efficiency.

Future development efforts will focus on these optimizations, particularly refining commutator handling and incorporating advanced computational techniques. Another key area of expansion involves extending SymPT to accommodate systems with an infinite number of subspaces. This feature would be invaluable for addressing many-body quantum problems, such as those encountered in the Anderson impurity model~\cite{SW_original_paper_on_SW}. Progress in this direction is underway, with ongoing work on integrating Einstein summation conventions into the software, which would streamline the treatment of multi-subspace systems.

In summary, SymPT represents a significant step forward in the automation and application of the SWT, providing a robust and adaptable platform for quantum research. We hope that by addressing its current limitations and expanding its capabilities, the software will have the potential to further aid researchers in a variety of fields, enabling deeper insights into complex quantum phenomena.

\section*{Acknowledgements}
We acknowledge interesting discussions with Benjamin D'Anjou regarding future problems to be tackled with SymPT. In addition, we sincerely thank Adam Asaad for his invaluable initial consultations, which have left a lasting impact on the implementation of this project. 
GFD and MB acknowledge funding from the Emmy Noether Programme of the German Research Foundation (DFG) under grant no. BE 7683/1-1.
LR and MB acknowledge financial support from the University of Augsburg through seed funding project 2023-26.

\paragraph{Author contributions}


\begin{appendix}
\section{Derivation of Perturbative Terms} \label{sec: multi-block generator}
In this section we derive a general form for the antihermitian operator $S^{(j)}$ (of order $j$) generating the perturbative block-diagonalization transformation under least action conditions (see Sec.~\ref{sec: multi-block introduction} for additional details). To begin, note that any operator $A = e^{\pm \Theta}$, where $\Theta = \sum_{\theta=1}^\infty \Theta^{(\theta)}$, can be expanded as
\begin{equation}
    A = \sum_{n=0}^\infty \frac{(\pm 1)^n}{n!} \Theta^n.
\end{equation}
Additionally, each power $\Theta^n$ can be further expanded as
\begin{equation}
    \Theta^n = \left(\sum_{\theta=1}^\infty \Theta^{(\theta)}\right)^n \equiv \sum_{\vec{\theta}_n} \Theta^{(\vec{\theta}_n)},
\end{equation}
where $\vec{\theta}_n = (\theta_1, \ldots, \theta_n)$ and $\Theta^{(\vec{\theta}_n)} \equiv \prod_{i=1}^n \Theta^{(\theta_i)}$. Note that the order of a term $\Theta^{(\vec{\theta})}$ is given by $\sum_{i=1}^n \theta_i$. Lastly, to classify terms systematically, we introduce the following definitions:

\begin{definition}[The set $\mathcal{T}(j, n)$]\label{def:Tan}
     This set organizes terms with total order $j$ and length $n = \text{dim}(\vec\theta)$, and is defined as  

\begin{equation}
    \mathcal{T}(j, n) \equiv \left\{ \vec{\theta} = (\theta_1, \dots, \theta_n) \,\middle|\, \theta_i \in \mathbb{Z}^+, \, \sum_{i=1}^n \theta_i = j \right\}.
\end{equation}

\end{definition}

\begin{definition}[The set $\mathcal{P}(j)$]\label{def:Pa}
    This set collects unique terms with total order $j$, irrespective of their length $n$, and is defined as  

\begin{equation}
    \mathcal{P}(j) \equiv \bigcup_{n=1}^j \mathcal{T}(j, n). \label{eq: partitions set P}
\end{equation}
\end{definition}

Using the definitions \ref{def:Tan} and \ref{def:Pa} , the terms of $A$ can be grouped by their order $j$, facilitating a systematic expansion. Therefore, the expansion of $A$ can be rewritten as
\begin{equation}\label{eq:Aa}
    A = I + \sum_{j=1}^\infty A^{(j)}, \quad A^{(j)} = \sum_{\vec{\theta} \in \mathcal{P}(j)} \frac{(\pm 1)^{\mathrm{dim}(\vec{\theta})}}{\mathrm{dim}(\vec{\theta})!} \Theta^{(\vec{\theta})}.
\end{equation}

Similarly to~\cite{DiVicenzo} the method begins by performing a perturbative full diagonalization, expressing $X$ as $X = e^{-Z}$, where $Z = \sum_{j=1}^\infty Z^{(j)}$. The terms $Z^{(j)}$ are determined by the condition specified in Eq.~(\ref{eq: condition on full diagonalization SW}). Once $Z$ is known, the operators $S^{(j)}$ are calculated in terms of $X$.  Using the formalism derived in this section, the term $\mathcal{B}(X^\dagger) \mathcal{B}(X)$ in Eq.~(\ref{eq: multi-block unitary}) is then expanded as  
\begin{equation}
\mathcal{B}(X^\dagger) \mathcal{B}(X) = \left[I + \sum_{i=1}^\infty \sum_{\vec{\theta} \in \mathcal{P}(i)} \frac{1}{\mathrm{dim}(\vec{\theta})!} \mathcal{B}(Z^{(\vec{\theta})})\right] \left[I + \sum_{j=1}^\infty \sum_{\vec{\theta} \in \mathcal{P}(j)} \frac{(-1)^{\mathrm{dim}(\vec{\theta})}}{\mathrm{dim}(\vec{\theta})!} \mathcal{B}(Z^{(\vec{\theta})})\right],
\end{equation}
yielding  
\begin{equation}
\mathcal{B}(X^\dagger) \mathcal{B}(X) = I + \sum_{j=2}^\infty \varepsilon^{(j)},
\end{equation}
where
\begin{align}
\nonumber \varepsilon^{(i)} &\equiv \sum_{\substack{\vec{\theta} \in \mathcal{P}(i) \\ \mathrm{dim}(\vec{\theta})~\mathrm{even}}} \frac{2}{\mathrm{dim}(\vec{\theta})!} \mathcal{B}\left(Z^{(\vec{\theta})}\right)+ \\
&\quad + \sum_{(j,k) \in \mathcal{T}(i, 2)} \sum_{\substack{\vec{\theta} \in \mathcal{P}(j) \\ \vec{\phi} \in \mathcal{P}(k)}} \frac{(-1)^{\mathrm{dim}(\vec{\phi})}}{\mathrm{dim}(\vec{\theta})!~\mathrm{dim}(\vec{\phi})!} \mathcal{B}\left(Z^{(\vec{\theta})}\right)\mathcal{B}\left(Z^{(\vec{\phi})}\right).
\end{align}

Using then the series expansion $(1 + \varepsilon)^{-\frac{1}{2}} = 1 + \sum_{n=1}^\infty \binom{-\frac{1}{2}}{n} \varepsilon^n$,  
\begin{equation}
\left\{\mathcal{B}(X^\dagger) \mathcal{B}(X)\right\}^{-\frac{1}{2}} = I + \sum_{j=2}^{\infty
} \sum_{\vec{\theta} \in \mathcal{P}(j)} \binom{-\frac{1}{2}}{\mathrm{dim}(\vec{\theta})} \varepsilon^{(\vec{\theta})}.
\end{equation}
Similarly, $X^\dagger \mathcal{B}(X)$ expands as $X^\dagger \mathcal{B}(X) = I + \sum_{j=1}^\infty V^{(j)}$ with
\begin{align}
\nonumber V^{(i)} &= \sum_{\vec{\theta} \in \mathcal{P}(i)} \frac{1}{\mathrm{dim}(\vec{\theta})!} \left[ Z^{(\vec{\theta})} + (-1)^{\mathrm{dim}(\vec{\theta})} \mathcal{B}\left(Z^{(\vec{\theta})}\right) \right]+ \\
&\quad + \sum_{(j,k) \in \mathcal{T}(i, 2)} \sum_{\substack{\vec{\theta} \in \mathcal{P}(j) \\ \vec{\phi} \in \mathcal{P}(k)}} \frac{(-1)^{\mathrm{dim}(\vec{\phi})}}{\mathrm{dim}(\vec{\theta})!~\mathrm{dim}(\vec{\phi})!} Z^{(\vec{\theta})} \mathcal{B}\left(Z^{(\vec{\phi})}\right).
\end{align}
Combining these, the full expansion of $U^\dagger$ becomes  
\begin{equation}
U^\dagger = \left[I + \sum_{i=1}^\infty V^{(i)}\right] \left[I + \sum_{j=2}^\infty \sum_{\vec{\theta} \in \mathcal{P}(j)} \binom{-\frac{1}{2}}{\mathrm{dim}(\vec{\theta})} \varepsilon^{(\vec{\theta})}\right] = I + \sum_{j=1}^\infty U^{(j)},
\end{equation}
with
\begin{align}
U^{(i)} &= V^{(i)} + \sum_{\vec{\theta} \in \mathcal{P}(i)} \binom{-\frac{1}{2}}{\mathrm{dim}(\vec{\theta})} \varepsilon^{(\vec{\theta})}+ \sum_{(j,k) \in \mathcal{T}(i, 2)} \sum_{\vec{\theta} \in \mathcal{P}(k)} \binom{-\frac{1}{2}}{\mathrm{dim}(\vec{\theta})} V^{(j)} \varepsilon^{(\vec{\theta})}.
\end{align}
Moreover, the term $U^\dagger = e^S$ can also be expanded as in Eq.~(\ref{eq:Aa}), where
\begin{align}
U^{(j)} &= \sum_{\vec{\theta} \in \mathcal{P}(j)} \frac{1}{\mathrm{dim}(\vec{\theta})!} S^{(\vec{\theta})} \\
&= S^{(j)} + \sum_{\substack{\vec{\theta} \in \mathcal{P}(j) \\ \mathrm{dim}(\vec{\theta}) \neq 1}} \frac{1}{\mathrm{dim}(\vec{\theta})!} S^{(\vec{\theta})}.
\end{align}

Finally, the generator $S^{(j)}$ is obtained iteratively as  

\begin{equation}
S^{(j)} = U^{(j)} - \sum_{\substack{\vec{\theta} \in \mathcal{P}(j) \\ \mathrm{dim}(\vec{\theta}) \neq 1}} \frac{1}{\mathrm{dim}(\vec{\theta})!} S^{(\vec{\theta})}. \label{eq: multiblock generator}
\end{equation}  

This recursive structure allows $S^{(\tau)}$ to be determined at any order $\tau$, facilitating the derivation of high-order block diagonalization terms.

\section{Expanded corrections transmon-resonator example}\label{sec: transmon resonator corretions}
In this section we include the expanded form of the correction terms presented in Eq.~(\ref{eq: second order correction to transmon resonator})
\begin{align}
    \nonumber \Omega_t  = & \frac{2 g^{2}}{N_{t} \alpha - \alpha - \omega_{r} + \omega_{t}} - \frac{2 g^{2}}{N_{t} \alpha + \omega_{r} + \omega_{t}} + \frac{\alpha g^{2} + g^{2} \omega_{r} - g^{2} \omega_{t}}{\left(N_{t} \alpha - \alpha - \omega_{r} + \omega_{t}\right)^{2}}+ \\
    &+ \frac{\alpha g^{2} + g^{2} \omega_{r} + g^{2} \omega_{t}}{\left(N_{t} \alpha + \omega_{r} + \omega_{t}\right)^{2}},\\
    \Omega_r = & - \frac{2 g^{2}}{N_{t} \alpha + \omega_{r} + \omega_{t}} - \frac{2 g^{2}}{N_{t} \alpha - \omega_{r} + \omega_{t}} + \frac{- g^{2} \omega_{r} + g^{2} \omega_{t}}{\left(N_{t} \alpha - \omega_{r} + \omega_{t}\right)^{2}} + \frac{g^{2} \omega_{r} + g^{2} \omega_{t}}{\left(N_{t} \alpha + \omega_{r} + \omega_{t}\right)^{2}},\\
    \alpha' = & - \frac{\alpha g^{2}}{\left(N_{t} \alpha - \alpha - \omega_{r} + \omega_{t}\right)^{2}} + \frac{\alpha g^{2}}{\left(N_{t} \alpha + \omega_{r} + \omega_{t}\right)^{2}}
\end{align}

\end{appendix}




\bibstyle{SciPost_bibstyle} 
\bibliography{references}

\end{document}